\documentclass{amsart}  
\usepackage{amsmath, latexsym, amsthm, amsfonts,bm,amssymb,mathscinet} % Useful basic packages.
\usepackage{mathtools}
\usepackage{xcolor}
\usepackage[plainpages=false,pdfpagelabels]{hyperref}
 
\usepackage{url} % Package for embedding url.
\usepackage{enumerate}
\usepackage{units} % nicefrac
\usepackage[nobysame,msc-links,author-year,bibtex-style]{amsrefs} %has to be commented

\theoremstyle{plain}

\theoremstyle{remark}
\newtheorem{remark}{Remark}

\newcommand{\ind}{\mathbf{1}}

\def\D{{ {\mathrm{d}}}}

\renewcommand{\th}{\theta}

\newcommand{\ig}{\operatorname{IG}}

\usepackage[OT2,OT1]{fontenc}
\newcommand\cyr{%
\renewcommand\rmdefault{wncyr}%
\renewcommand\sfdefault{wncyss}%
\renewcommand\encodingdefault{OT2}%
\normalfont
\selectfont}
\DeclareTextFontCommand{\textcyr}{\cyr}

\newcommand{\loc}{./} % change when uploading to arxiv ./

\definecolor{violet}{rgb}{0.3,0.0, 0.55}

\def\epsilon{\varepsilon}

\allowdisplaybreaks

\begin{document}

\title[Nonparametric {B}ayesian volatility estimation]{Nonparametric {B}ayesian volatility estimation}

\author[S.~Gugushvili]{Shota Gugushvili}
\address{Biometris\\
	Wageningen University \& Research\\
	Postbus 16\\
	6700 AA Wageningen\\
	The Netherlands}
\email{shota@yesdatasolutions.com}
\thanks{The research leading to the results in this paper has received funding from the European Research Council under ERC Grant Agreement 320637.}

\author[F. H.~van der Meulen]{Frank van der Meulen}
\address{Delft Institute of Applied Mathematics\\
Faculty of Electrical Engineering, Mathematics and Computer Science\\
Delft University of Technology\\
Van Mourik Broekmanweg 6  \\
2628 XE Delft\\
The Netherlands}
\email{f.h.vandermeulen@tudelft.nl}

\author[M.~Schauer]{Moritz Schauer}
\address{Mathematical Institute\\
Leiden University\\
P.O. Box 9512\\
2300 RA Leiden\\
The Netherlands}
\email{m.r.schauer@math.leidenuniv.nl}

\author[P.~Spreij]{Peter Spreij} 
\address{Korteweg-de Vries Institute for Mathematics\\
University of Amsterdam\\
P.O. Box 94248\\
1090 GE Amsterdam\\
The Netherlands \and Institute for Mathematics, Astrophysics and Particle Physics\\ Radboud University\\ Nijmegen\\ The Netherlands}
\email{spreij@uva.nl}



\begin{abstract}

Given discrete time observations over a fixed time interval, we study a nonparametric Bayesian approach to estimation of the volatility coefficient of a stochastic differential equation. We postulate a histogram-type prior on the volatility with piecewise constant realisations on bins forming a partition of the time interval. The values on the bins are assigned  an inverse Gamma Markov chain (IGMC) prior. Posterior inference is straightforward to implement via Gibbs sampling, as the full conditional distributions are available explicitly and turn out to be inverse Gamma. We also discuss in detail the hyperparameter selection for our method. Our nonparametric Bayesian approach leads to good practical results in representative simulation examples. Finally, we apply it on a classical data set in change-point analysis: weekly closings of the Dow-Jones industrial averages.

\end{abstract}

\date{\today}

\maketitle

%%%%%%%%%%%%%%%%
\section{Introduction}

\subsection{Problem formulation} Consider a one-dimensional stochastic differential equation (SDE)
\begin{equation}
\label{sde}
\D X_t=b_0(t,X_t)\,\D  t+s_0(t)\,\D  W_t, \quad X_0=x, \quad t\in[0,T],
\end{equation}
where $b_0$ is the drift coefficient, $s_0$ the deterministic dispersion coefficient or volatility, and $x$ is a deterministic initial condition. Here $W$ is a standard Brownian motion. Assume that standard conditions for existence and uniqueness of a strong solution to \eqref{sde} are satisfied (see, e.g., \cite{karatzas91}), and observations
\[
\scr{X}_n =\{X_{t_{0,n}}, \ldots, X_{t_{n,n}}\}
\]
are available, where $t_{i,n}=iT/n,$ $i=0,\ldots,n.$ Using a nonparametric Bayesian approach, our aim is to estimate the volatility function $s_0$. In a financial context, knowledge of the volatility is of fundamental importance e.g.\ in pricing financial derivatives; see \cite{bjork09} and \cite{musiela05}. However, SDEs have applications far beyond the financial context as well, e.g.\ in physics, biology, life sciences, neuroscience and engineering (see \cite{allen07}, \cite{fuchs13}, \cite{hindriks11} and \cite{wong85}). Note that by It\^{o}'s formula, using a simple transformation of the state variable, also an SDE of the form
\[
\D  X_t=b_0(t,X_t)\,\D  t+s_0(t)f_0(X_t)\,\D  W_t, \quad X_0=x, \quad t\in[0,T],
\]
can be reduced to the form \eqref{sde}, provided the function $f_0$ is known and regular enough; see, e.g., p.~186 in \cite{soulier98}. Some classical examples that fall under our statistical framework are the geometric Brownian motion and the Ornstein-Uhlenbeck process. Note also that as we allow the drift in \eqref{sde} to be non-linear, marginal distributions of $X$ are not necessarily Gaussian and may thus exhibit heavy tails, which is attractive in financial modelling.

A nonparametric approach guards one against model misspecification and is an excellent tool for a preliminary, exploratory data analysis, see, e.g., \cite{silverman86}. Commonly acknowledged advantages of a Bayesian approach include automatic uncertainty quantification in parameter estimates via Bayesian credible sets, and the fact that it is a fundamentally likelihood-based method. In \cite{mueller13} it has been argued that a nonparametric Bayesian approach is important for honest representation of uncertainties in inferential conclusions. Furthermore, use of a prior allows one to easily incorporate the available external, a priori information into the estimation procedure, which is not straightforward to achieve with frequentist approaches. For instance, this a priori information could be an increasing or decreasing trend in the volatility.

\subsection{Literature overview} Literature on nonparametric Bayesian volatility estimation in SDE models is scarce. We can list theoretical contributions \cite{gugu14}, \cite{gugu16}, \cite{nickl17b}, and the practically oriented paper \cite{batz17}. The model in the former two papers is close to the one considered in the present work, but from the methodological point of view different Bayesian priors are used and practical usefulness of the corresponding Bayesian approaches is limited. On the other hand, the models considered in \cite{nickl17b} and \cite{batz17} are rather different from ours, and so are the corresponding Bayesian approaches. The nearest predecessor of the model and the method in our paper is the one studied in \cite{gugu17}. In the sequel we will explain in what aspects the present contribution differs from that one and what the current improvements are. We note in passing that there exists a solid body of literature on nonparametric Bayesian estimation of the drift coefficient, see, e.g., \cite{gugu14b}, \cite{papaspiliopoulos12}, \cite{pokern13}, \cite{ruttor13}, \cite{frank13}, \cite{frank14} and the review article \cite{harry13}, but Bayesian volatility estimation requires use of substantially different ideas. We also note existence of works dealing with parametric Bayesian estimation in discrete-time stochastic volatility models, see, e.g., \cite{jacquier94} and \cite{jacquier04}, but again, these are not directly related to the problem we study in this paper.

\subsection{Approach and results} The main potential difficulties facing a Bayesian approach to inference in SDE models from discrete observations are an intractable likelihood and absence of a closed form expression for the posterior distribution; see, e.g., \cite{elerian01},  \cite{fuchs13}, \cite{roberts01} and \cite{vdm-s-estpaper}. Typically, these difficulties necessitate the use of a data augmentation device (see \cite{tanner87}) and some intricate form of a Markov chain Monte Carlo (MCMC) sampler (see \cite{robert04}). In \cite{gugu17}, these difficulties are circumvented by intentionally setting the drift coefficient to zero, and employing a (conjugate) histogram-type prior on the diffusion coefficient, that has piecewise constant realisations on bins forming a partition of $[0,T]$. Specifically, the (squared) volatility is modelled a priori as a function
$
s^2=\sum_{k=1}^{N} \theta_k \ind_{B_k},
$
with independent and identically distributed inverse gamma coefficients $\theta_k$'s, and the prior $\Pi$ is defined as the law of $s^2$. Here $B_1,\ldots,B_N$ are bins forming a partition of $[0,T]$. With this independent inverse Gamma (IIG) prior, $\th_1,\ldots, \th_N$ are independent, conditional on the data, and of inverse gamma type. Therefore,  this approach results in a fast and simple to understand and implement Bayesian procedure. A study of its favourable practical performance, as well as its theoretical validation was recently undertaken in \cite{gugu17}. As shown there under precise regularity conditions, misspecification of the drift is asymptotically, as the sample size $n\rightarrow\infty$, harmless for consistent estimation of the volatility coefficient.

Despite a good practical performance of the method in \cite{gugu17}, there are some limitations associated with it too. Thus, the method offers limited possibilities for adaptation to the local structure of the volatility coefficient, which may become an issue if the volatility has a wildly varying curvature on the time interval $[0,T]$. A possible fix to this would be to equip the number of bins $N$ forming a partition of $[0,T]$ with a prior, and choose the endpoints of bins $B_k$ also according to a prior. However, this would force one to go beyond the conjugate Bayesian setting as in \cite{gugu17}, and  posterior inference in practice would require, for instance, the use of a reversible jump MCMC algorithm (see \cite{green95}). Even in the incomparably simpler setting of intensity function estimation for nonhomogeneous Poisson processes with histogram-type priors, this is very challenging, as observed in \cite{yang01}. Principal difficulties include designing moves between models of differing dimensions that result in MCMC algorithms that \emph{mix well}, and \emph{assessment of convergence} of Markov chains (see \cite{fearnhead06}, p.~204). Thus, e.g., the inferential conclusions in \cite{green95} and \cite{green03} are different on the same real data example using the same reversible jump method, since it turned out that in the first paper the chain was not run long enough. Cf.~also the remarks on Bayesian histograms in \cite{gelman:book:14}, p.~546.

Here we propose an alternative approach, inspired by ideas in \cite{cemgil07} in the context of audio signal modelling different from the SDE setting that we consider; see also  \cite{cemgil:proc:08}, \cite{cemgil07b}, \cite{dikmen08}, \cite{dikmen10} and \cite{virtanen08}. Namely, instead of using a prior on the (squared) volatility that has piecewise constant realisations on $[0,T]$ with independent coefficients $\theta_k$'s, we will assume that the sequence $\{\theta_k\}$ forms a suitably defined Markov chain. An immediately apparent advantage of using such an approach is that it induces extra smoothing via dependence in prior realisations of the volatility function across different bins.  Arguing heuristically, with a large number $N$ of bins $B_k$ it is then possible to closely mimick the local structure of the volatility: in those parts of the interval $[0,T]$, where the volatility has a high curvature or is subject to abrupt changes, a large number of (narrow) bins is required to adequately capture these features. However, the grid used to define the bins $B_k$'s is uniform, and if $\th_1,\ldots,\th_N$ are a priori independent, a large $N$ may induce spurious variability in the volatility estimates in those regions of $[0,T]$ where the volatility in fact varies slowly. As we will see in the sequel, this problem may be alleviated using a priori dependent $\theta_k$'s.

In the subsequent sections we detail our approach, and study its practical performance via simulation and real data examples. Specifically, we implement our method via a straightforward version of the Gibbs sampler, employing the fact that full conditional distributions of $\theta_k$'s are known in closed form (and are in fact inverse gamma). Unlike \cite{gugu17}, posterior inference in our new approach requires the use of MCMC. However, this is offset by the advantages of our new approach outlined above, and in fact the additional computational complexity of our new method is modest in comparison to \cite{gugu17}. The prior in our new method depends on hyperparameters, and we will also discuss several ways of their choice in practice.

\subsection{Organisation of this paper}  In Section~\ref{section:model} we supply a detailed description of our nonparametric Bayesian approach to volatility estimation. In Section~\ref{section:simulation} we study the performance of our method via extensive simulation examples. In Section~\ref{section:realdata} we apply the method on a real data example. Section~\ref{section:conclusions} summarises our findings and provides an outlook on our results. Finally, Section~\ref{section:details} contains some additional technical details of our procedure.

\subsection{Notation}
We denote the prior distribution on the (squared) volatility function by $\Pi$ and write the posterior measure given data $\mathcal{X}_n$ as $\Pi(\,\cdot\mid \mathcal{X}_n)$. 
We use the notation $\ig(\alpha,\beta)$ for the inverse gamma distribution with shape parameter $\alpha>0$ and scale parameter $\beta>0$. This distribution has a density
\begin{equation}
\label{ig:dens}
x \mapsto \frac{\beta^{\alpha}}{\Gamma(\alpha)} x^{-\alpha-1} e^{-\beta/x}, \quad x>0.
\end{equation}
For two sequences $\{a_n\}$, $\{b_n\}$, the notation $a_n \asymp b_n$ will be used to denote the fact that the sequences are asymptotically (as $n\rightarrow\infty$) of the same order. Finally, for a density $f$ and a function $g$, the notation $f \propto g $ will mean that $f$ is proportional to $g$, with proportionality constant on the righthand side recovered as $(\int g)^{-1}$, where the integral is over the domain of definition of $g$ (and of $f$). The function $g$ can be referred to as an unnormalised probability density.

%%%%%%%%%%%%%%%%
\section{Nonparametric Bayesian approach}\label{section:model}

\subsection{Generalities}

Our starting point is the same as in \cite{gugu17}. Namely, we misspecify the drift coefficient $b_0$ by intentionally setting it to zero (see also \cite{martin16} for a similar idea of `misspecification on purpose'). The theoretical justification for this under the `infill' asymptotics, with the time horizon $T$ staying fixed and the observation times $t_{i,n}=iT/n,$ $i=1,\ldots,n,$ filling up the interval $[0,T]$ as $n\rightarrow\infty$, is provided in \cite{gugu17}, to which we refer for further details (the argument there ultimately relies on Girsanov's theorem). Similar ideas are also encountered in the non-Bayesian setting in the econometrics literature on high-frequency financial data, see, e.g., \cite{mykland12}.

Set $Y_{i,n}=X_{t_{i,n}}-X_{t_{i-1,n}}$. With the assumption $b_0=0$, the pseudo-likelihood of our observations is tractable, in fact Gaussian,
\begin{equation}
\label{likelih}
L_n(s^2)=\prod_{i=1}^{n} \left\{ \frac{1}{\sqrt{2\pi\int_{t_{i-1,n}}^{t_{i,n}}s^2(u)\,\D  u}}\psi\left( \frac{Y_{i,n}}{\sqrt{\int_{t_{i-1,n}}^{t_{i,n}}s^2(u)\,\D  u}} \right) \right\},
\end{equation}
where $\psi(u)=\exp(-u^2/2).$ The posterior probability of any measurable set $S$ of volatility functions can be computed via Bayes' theorem as
\begin{equation*}
\Pi(S\mid \mathcal{X}_n)=\frac{\int_{S} L_n(s^2) \Pi(\D  s)}{ \int L_n(s^2) \Pi(\D  s) }.
\end{equation*}
Here the denominator is the normalising constant, the integral over the whole space on which the prior $\Pi$ is defined, which ensures that the posterior is a \emph{probability} measure (i.e.\ integrates to one).

\subsection{Prior construction}

Our prior $\Pi$ is constructed similarly to \cite{gugu17}, with an important difference to be noted below. Fix an integer $m<n$. Then $n=mN+r$  with $0\leq r<m$, where $N=\lfloor \frac{n}{m}\rfloor$. Now define bins $B_k=[t_{m(k-1),n},t_{mk,n})$, $k=1,\ldots,N-1$, and $B_N=[t_{m(N-1),n},T]$. Thus the first $N-1$ bins are of length $mT/n$, whereas the last bin $B_N$ has length $T-t_{m(N-1),n}=n^{-1}(r+m)T<n^{-1} 2mT$. The parameter $N$ (equivalently, $m$) is a hyperparameter of our prior. We model $s$ as piecewise constant on bins $B_k$, thus
$
s=\sum_{k=1}^{N} \xi_k \ind_{B_k}.
$
The prior $\Pi$ on the volatility $s$ can now be defined by assigning a prior to the coefficients $\xi_k$'s.

Let  $\theta_k=\xi_k^2$. Since the bins $B_k$ are disjoint,
\begin{equation}
\label{eq:s2}
s^2=\sum_{k=1}^{N} \xi_k^2 \ind_{B_k}=\sum_{k=1}^{N} \theta_k \ind_{B_k}.
\end{equation}
As the likelihood depends on $s$ only through its square $s^2$, it suffices to assign the prior to the coefficients $\theta_k$'s of $s^2$. This is the point where we fundamentally diverge from \cite{gugu17}. Whereas in \cite{gugu17} it is assumed that $\{\theta_k\}$ is an i.i.d.\ sequence of inverse gamma random variables, here we suppose that $\{\theta_k\}$ forms a Markov chain. This will be referred to as an inverse Gamma Markov chain (IGMC) prior (see \cite{cemgil07}), and is defined as follows. Introduce auxiliary variables $\zeta_k,$ $k=2,\ldots,N$, and define a Markov chain using the time ordering $\theta_1,\zeta_2,\theta_2,  \ldots,\zeta_k,\theta_k,\ldots,\zeta_N,\theta_N$. Transition distributions of this chain are defined as follows: fix hyperparameters $\alpha_1$, $\alpha_{\zeta}$ and $\alpha$, and set
\begin{equation}
\label{formula:prior}
\theta_1 \sim \ig(\alpha_{1},\alpha_{1}), \quad \zeta_{k+1} | \theta_k \sim \ig(\alpha_{\zeta},\alpha_{\zeta} \theta_k^{-1}), \quad \theta_{k+1} | \zeta_{k+1} \sim \ig(\alpha,\alpha \zeta_{k+1}^{-1}).
\end{equation}
The name of the chain reflects the fact that these distributions are inverse Gamma.

\begin{remark}
Our definition of the IGMC prior differs from the one in  \cite{cemgil07} in the choice of the initial distribution of $\theta_1$, which is important to alleviate possible `edge effects' in volatility estimates in a neighbourhood of $t=0$. The parameter $\alpha_1$ determines the initial distribution of the inverse Gamma Markov chain. Letting $\alpha_1 \to 0$ (which corresponds to a vague prior) `releases' the chain at the time origin. \qed
\end{remark}

\begin{remark}
As observed in \cite{cemgil07}, there are various ways of defining an inverse Gamma Markov chain. The point to be kept in mind is that the resulting posterior should be computationally tractable, and the prior on $\theta_k$'s should have a capability of producing realisations with positive correlation structures, as this introduces smoothing among the $\theta_k$'s in adjacent bins. This latter property is not possible to attain with arbitrary constructions of inverse Gamma Markov chains, such as e.g.\ a natural construction $\theta_k|\theta_{k-1}\sim \ig(\alpha,\theta_{k-1}/\alpha)$. On the other hand, positive correlation between realisations $\theta_k$'s can be achieved e.g. by setting $\theta_k | \theta_{k-1} \sim \ig(\alpha,(\alpha \theta_{k-1})^{-1})$, but this results in intractable posterior computations. The definition of the IGMC prior in the present work, that employs latent variables $\zeta_k$'s, takes care of both these important points. For an additional discussion see \cite{cemgil07}. \qed
\end{remark}

\begin{remark}
\label{rem:dependence}
Setting the drift coefficient $b_0$ to zero effectively results in pretending that the process $X$ has independent (Gaussian) increments. In reality, since the drift in practical applications is typically nonzero, increments of the process are dependent, and hence all observations $Y_{i,n}$ contain some indirect information on the value of the volatility $s^2$ at each time point $t\in[0,T]$. On the other hand, assuming the IGMC prior on $s^2$ yields a posteriori dependence of coefficients $\{\theta_k\}$, which should be of help in inference with smaller sample sizes $n$. See Section~\ref{section:realdata} for an illustration. \qed 
\end{remark}

\subsection{Gibbs sampler}

\begin{figure}
\includegraphics[width=0.32\textwidth]{\loc 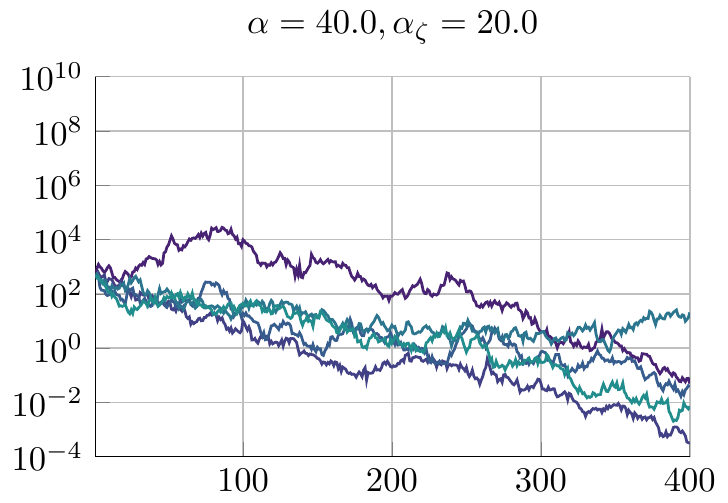}
\includegraphics[width=0.32\textwidth]{\loc 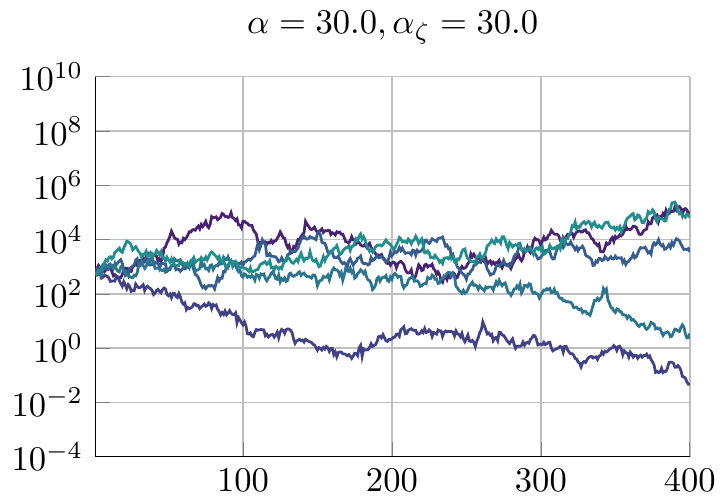}
\includegraphics[width=0.32\textwidth]{\loc 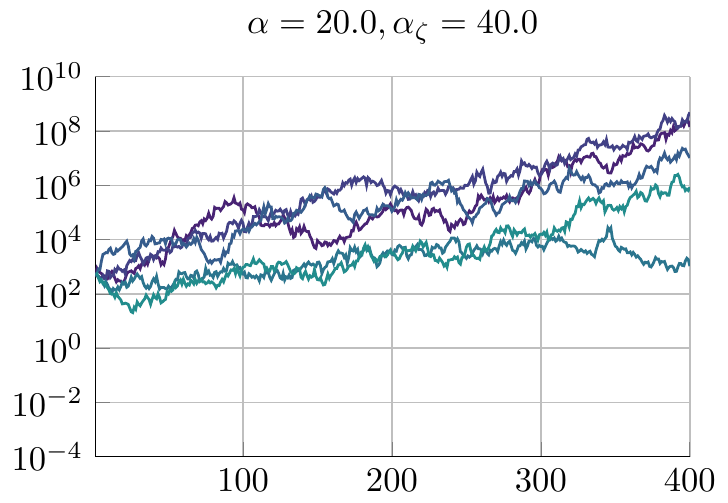}
\caption{Realisations of the Markov chain $\{\theta_k\}$ with $\alpha=40,$ $\alpha_{\zeta}=20$ (left panel) and $\alpha=30,$ $\alpha_{\zeta}=30$ (center panel) and $\alpha=20,$ $\alpha_{\zeta}=40$ (right panel). In all cases, $\theta_1$ is fixed to $500$ .}
\label{fig:prior}
\end{figure}

It can be verified by direct computations employing \eqref{formula:prior} that the full conditional distributions of $\theta_k$'s and $\zeta_k$'s are inverse gamma,
\begin{align}
\theta_k | \zeta_k,\zeta_{k+1} &\sim \ig\left(\alpha+\alpha_{\zeta},\frac{\alpha}{\zeta_k}+\frac{\alpha_{\zeta}}{ \zeta_{k+1}}\right), \quad k=2,\ldots,N-1, \label{fullcond1} \\
\theta_1 | \zeta_2 & \sim \ig\left(\alpha_1+\alpha_{\zeta},  \alpha_1 + \frac{\alpha_{\zeta}}{ \zeta_{2}} \right), \label{fullcondstart} \\
\theta_N | \zeta_N & \sim \ig\left(\alpha,\frac{\alpha}{\zeta_N}\right), \label{fullcondend} \\
\zeta_k | \theta_k,\theta_{k-1} &\sim \ig\left(\alpha_{\zeta}+\alpha, \frac{\alpha_{\zeta}}{\theta_{k-1}}+\frac{\alpha}{ \theta_k}\right), \quad k=2,\ldots,N. \label{fullcond2}
\end{align}
See Section \ref{section:details} for details. Next, the effective transition kernel of the Markov chain $\{\theta_k\}$ is given by formula (4) in \cite{cemgil07}, and is a scale mixture of inverse gamma distributions; however, its exact expression is of no direct concern for our purposes. As noted in \cite{cemgil07}, p.~700, depending on the parameter values $\alpha,\alpha_{\zeta}$, it is possible for the chain $\{\theta_k\}$ to exhibit either an increasing or decreasing trend. We illustrate this point by plotting realisations of $\{\theta_k\}$ in Figure~\ref{fig:prior} for different values of $\alpha$ and $\alpha_{\zeta}$. In the context of volatility estimation this feature is attractive, if prior information on the volatility trend is available.

Inference in \cite{cemgil07} is performed using a mean-field variational Bayes approach, see, e.g.,  \cite{blei16}. Here we describe instead a fully Bayesian approach relying on Gibbs sampling (see, e.g.,  \cite{gelfand90} and \cite{geman84}), cf.~\cite{cemgil07b}.

The algorithm is initialised at values $\zeta_2,\ldots,\zeta_N$, e.g.\ generated from the prior specification \eqref{formula:prior}. In order to derive update formulae for the full conditionals of the $\theta_k$'s, define
\begin{align*}
Z_k &= \sum_{i=(k-1)m+1}^{km} Y_{i,n}^2, \quad k=1,\ldots,N-1,\\
Z_N &= \sum_{i=(N-1)m+1}^{n} Y_{i,n}^2.
\end{align*}
With this notation, the likelihood from \eqref{likelih} satisfies
\begin{equation}
\label{eq:likelih}
L_n(\th) \propto  \theta_N^{-(m+r)/2} \exp\left(-\frac{n Z_N}{2T\th_N} \right) \prod_{k=1}^{N-1} \th_k^{-m/2} \exp\left(-\frac{n Z_k}{2T\th_k} \right).
\end{equation}
Using this formula and equation \eqref{fullcond1}, and recalling the form of the inverse gamma density \eqref{ig:dens}, it is seen that the update distribution for $\theta_k,$ $k=2,\ldots,N-1$, is
\[
\mathrm{IG}\left(\alpha+\alpha_{\zeta}+\frac{m}{2},\, \frac{\alpha}{\zeta_k}+\frac{\alpha_{\zeta}}{\zeta_{k+1}} + \frac{n Z_k}{2T}\right),
\]
whereas by \eqref{fullcondend} the ones for $\theta_1$ and $\theta_N$ are
\[
\mathrm{IG}\left(\alpha_1 + \alpha_\zeta +\frac{m}{2},\, \alpha_1 +  \frac{\alpha_\zeta}{ \zeta_{2}} + \frac{n Z_1}{2T}\right),\quad \mathrm{IG}\left(\alpha+\frac{m+r}{2},\, \frac{\alpha}{ \zeta_N} + \frac{n Z_N}{2T}\right),
\]
respectively.

Next, the latent variables $\zeta_k$'s will be updated using formula \eqref{fullcond2}. This update step for $\zeta_k$'s does not directly involve the data $\mathcal{X}_n$, except through the previous values of $\theta_k$'s.

Finally, one iterates these two Gibbs steps for $\theta_k$'s and $\zeta_k$'s a large number of times (until chains can be assessed as reasonably converged), which gives posterior samples of the $\theta_k$'s. Using the latter, the posterior inference can proceed in the usual way, e.g.\ by computing the sample posterior mean of $\theta_k$'s, as well as sample quantiles, that provide, respectively, a point estimate and uncertainty quantification via marginal Bayesian credible bands for the squared volatility $s^2$. Similar calculations on the square roots of the posterior samples can be used to obtain point estimates and credible bands for the volatility function $s$ itself.

\subsection{Hyperparameter choice}

We first assume the number of bins $N$ has been chosen in some way, and we only have to deal with hyperparameters $\alpha$, $\alpha_{\zeta}$ and $\alpha_1$, that govern properties of the Markov chain prior. In \cite{cemgil07}, where an IGMC prior was introduced, guidance on the hyperparameter selection is not discussed. In \cite{cemgil:proc:08}, the hyperparameters are fine-tuned by hand in specific problems studied there (audio denoising and single channel audio source separation). Another practical solution is to try several different fixed combinations of the hyperparameters $\alpha$, $\alpha_{\zeta}$ and $\alpha_1$, if only to verify sensitivity of inferential conclusions with respect to variations in the hyperparameters. Some further methods for hyperparameter optimisation are discussed in \cite{dikmen10}. In \cite{cemgil:proc:08} optimisation of the hyperparameters via the maximum likelihood method is suggested; practical implementation relies on the EM algorithm (see \cite{dempster77}), and some additional details are given in \cite{dikmen08}. Put in other terms, the proposal in \cite{dikmen08} amounts to using an empirical Bayes method (see, e.g.,  \cite{efron10}, \cite{robbins56} and \cite{robbins64}). The use of the latter is widespread and often leads to good practical results, but the method is still insufficiently understood theoretically, except in toy models like the white noise model (see, however, \cite{donnet18} and \cite{petrone14} for some results in other contexts). On the practical side, in our case, given that the dimension of the sequences $\{\zeta_k\}$ and $\{\theta_k\}$ is rather high, namely $2N-1$ with $N$ large, and the marginal likelihood is not available in closed form, this approach is expected to be computationally intensive. Therefore, a priori there is no reason not to try instead a fully  Bayesian approach by equiping the hyperparameters with a prior, and this is in fact our default approach in the present work. However, the corresponding full conditional distribution turns out to be nonstandard, which necessitates the use of a Metropolis-Hastings step within the Gibbs sampler (see, e.g., \cite{hastings70}, \cite{metropolis53} and \cite{tierney94}). We provide the necessary details in Section~\ref{section:details}.

Finally, we briefly discuss the choice of the hyperparameter $N$. As argued in \cite{gugu17}, in practice it is recommended to use the theoretical results in \cite{gugu17} (that suggest to take $N \asymp n^{\lambda/(2\lambda+1)}$, if the true volatility function $s_0$ is $\lambda$-H\"{o}lder smooth) and try several values of $N$ simultaneously. Different $N$'s all provide information on the unknown volatility, but at different resolution levels; see Section 5 in \cite{gugu17} for an additional discussion. As we will see in simulation examples in Section \ref{section:simulation}, inferential conclusions with the IGMC prior are quite robust with respect to the choice of $N$. This is because  through the hyperparameters $\alpha$ and $\alpha_{\zeta}$, the IGMC prior has an additional layer for controlling the amount of applied smoothness; when $\alpha$ and $\alpha_{\zeta}$ are equipped with a prior (as above), they can in fact be learned from the data.

%%%%%%%%%%%%%%%%
\section{Synthetic data examples}
\label{section:simulation}

Computations in this section have been done in the programming language Julia, see \cite{bezanson17}. In order to test the practical performance of our estimation method, we use a challenging example with  the blocks function from \cite{donoho95}. As a second example, we consider the case of the Cox-Ross-Ingersoll model. Precise details are given in the subsections below.

We used the Euler scheme on a grid with 800\,001 equidistant points on the interval $[0,1]$ to obtain realisations of a solution to \eqref{sde} for different combinations of the drift and dispersion coefficients. These were then subsampled to obtain $n=4000$ observations in each example.

The hyperparameter $\alpha_1$ was set to $0.1$, whereas for the other two hyperparameters we assumed that $\alpha=\alpha_{\zeta}$ and used a diffuse $\ig(0.3,0.3)$ prior, except in specially noted cases below. Inference was performed using the Gibbs sampler from Section~\ref{section:model}, with a Metropolis-Hastings step to update the hyperparameter $\alpha$. The latter used an independent Gaussian random walk proposal with a scaling to ensure the acceptance rate of ca.~$50\%$; see Section~\ref{section:details}. The Gibbs sampler was run for $200 \, 000$ iterations and we used a burn-in of $1000$ samples. In each example we plotted $95\%$ marginal credible bands obtained from the central posterior intervals for the coefficients $\xi_k=\sqrt{\theta}_k$.

\subsection{Blocks function}

As our first example, we considered the case when the volatility function was given by the blocks function from \cite{donoho95}. With a vertical shift for positivity, this is defined as follows:
\begin{equation}\label{s3}
s(t)= 10 + 3.655606\times\sum_{j=1}^{11} h_j K(t-t_j), \quad t\in[0,1],
\end{equation}
where $K(t)=(1+\operatorname{sgn}(t))/2$, and
\begin{align*}
\{t_j\}&=(0.1,0.13,0.15,0.23,0.25,0.4,0.44,0.65,0.76,0.78,0.81),\\
\{h_j\}&=(4,-5,3,-4,5,-4.2,2.1,4.3,-3.1,2.1,-4.2).
\end{align*}
The function serves as a challenging benchmark example in nonparametric regression: it is mostly very smooth, but spatially inhomogeneous and characterised by abrupt changes (cf.~Chapter~9 in \cite{wasserman06}). Unlike nonparametric regression, the noise (Wiener process) in our setting should be thought of as multiplicative and proportional to $s$ rather than additive, which combined with the fact that $s$ takes rather large values further complicates the inference problem. Our main goal here was to compare the performance of the IGMC prior-based approach to the IIG prior-based one from \cite{gugu17}. To complete the SDE specification, our drift coefficient was chosen to be a rather strong linear drift $b_0(x) = -10x+20.$ 

\begin{figure}
\includegraphics[width=0.48\textwidth]{\loc 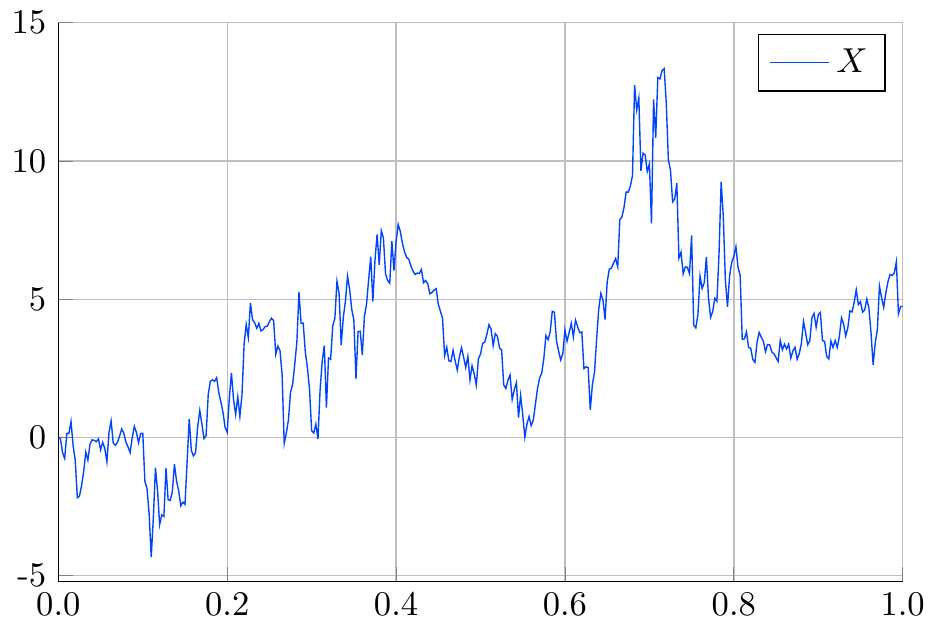}
\includegraphics[width=0.48\textwidth]{\loc 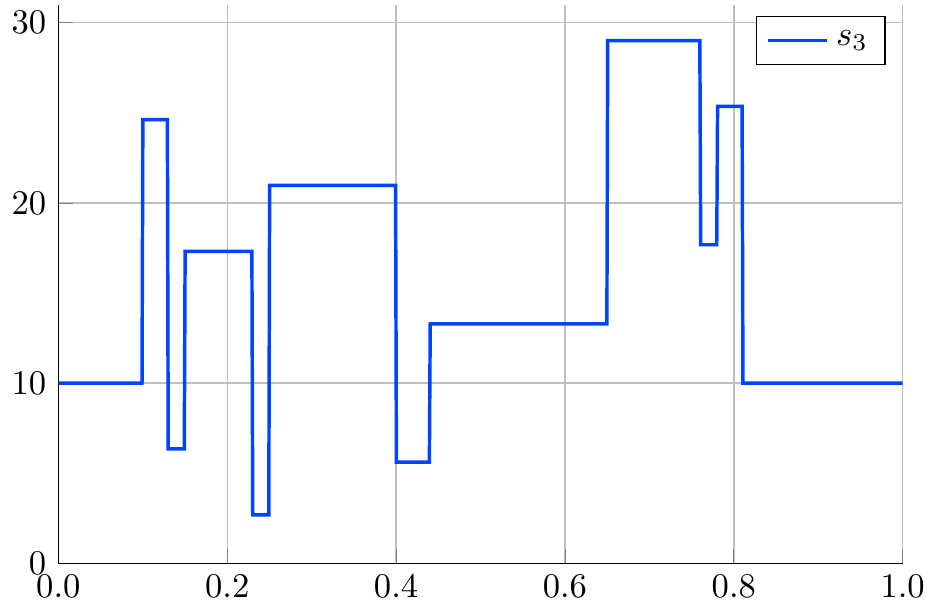}
\caption{The  sample path of the process $X$ from \eqref{s3} (left panel) and the corresponding volatility function $s$  (right panel).}
\label{fig:blockspath}
\end{figure}

In Figure~\ref{fig:blockspath} we plot the blocks function \eqref{s3} and the corresponding realisation of the process $X$ used in this simulation run.

\begin{figure}
\includegraphics[width=0.48\textwidth]{\loc 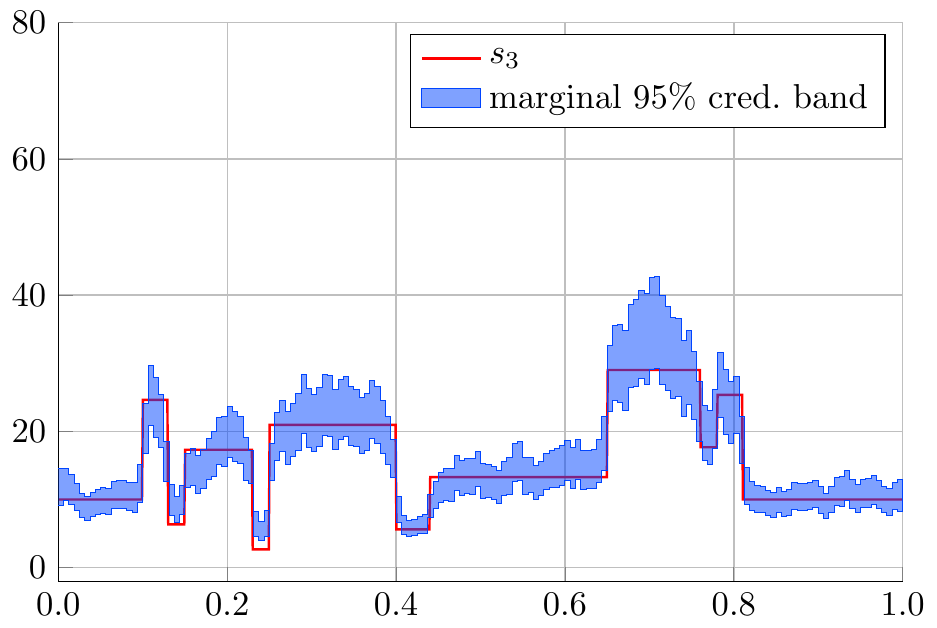}
\includegraphics[width=0.48\textwidth]{\loc 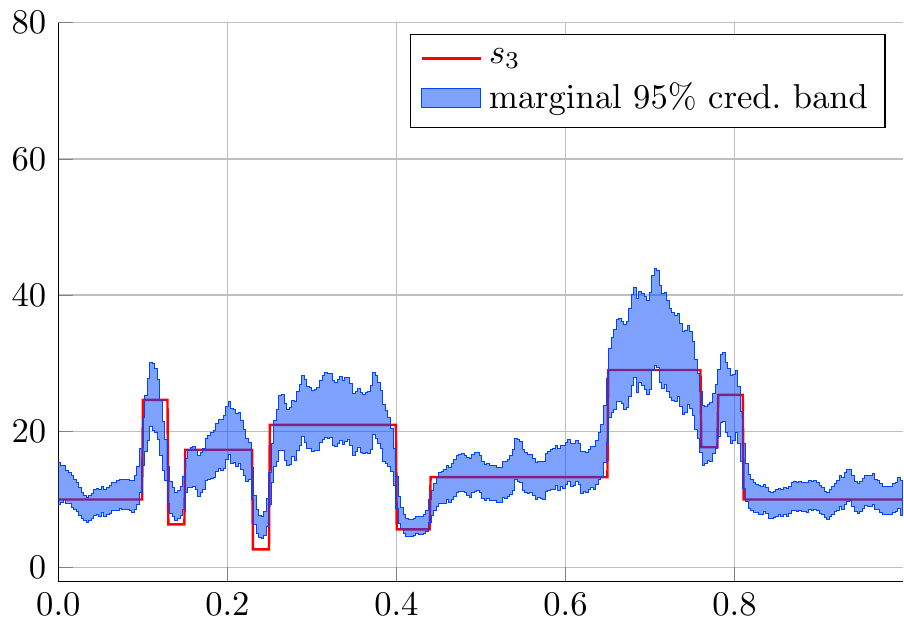}
\caption{Volatility function $s$ from \eqref{s3} with superimposed $95\%$ marginal credible band for the IGMC prior, using $N=160$, $\alpha =\alpha_{\zeta} = 20$ (left panel) and  $N=320$, $\alpha =\alpha_{\zeta} = 40$ (right panel); in both cases, $\alpha_1=0.1$.}
\label{fig:blocks}
\end{figure}

The left and right panels of Figure~\ref{fig:blocks} contrast the results obtained using the IGMC prior with $N=160$ and $\alpha=\alpha_{\zeta}=20$ versus $N=320$ and $\alpha=\alpha_{\zeta}=40$. These plots illustrate the fact that increasing $N$ has the effect of undersmoothing prior realisations, that can be balanced by increasing the values of $\alpha_{\zeta},\alpha$, which has the opposite smoothing effect. Because of this, in fact, both plots look quite similar.

\begin{figure}
\includegraphics[width=0.48\textwidth]{\loc 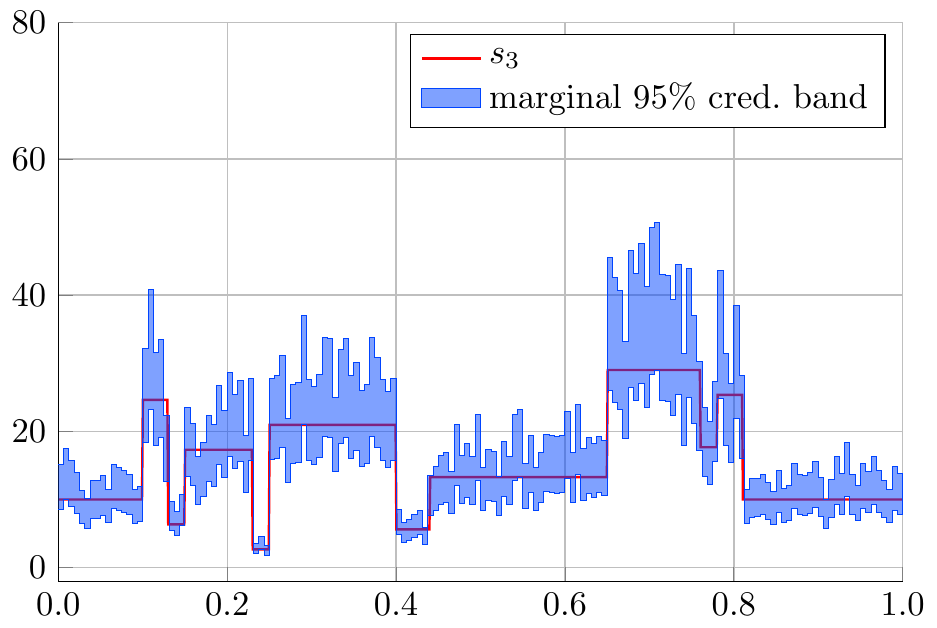}
\includegraphics[width=0.48\textwidth]{\loc 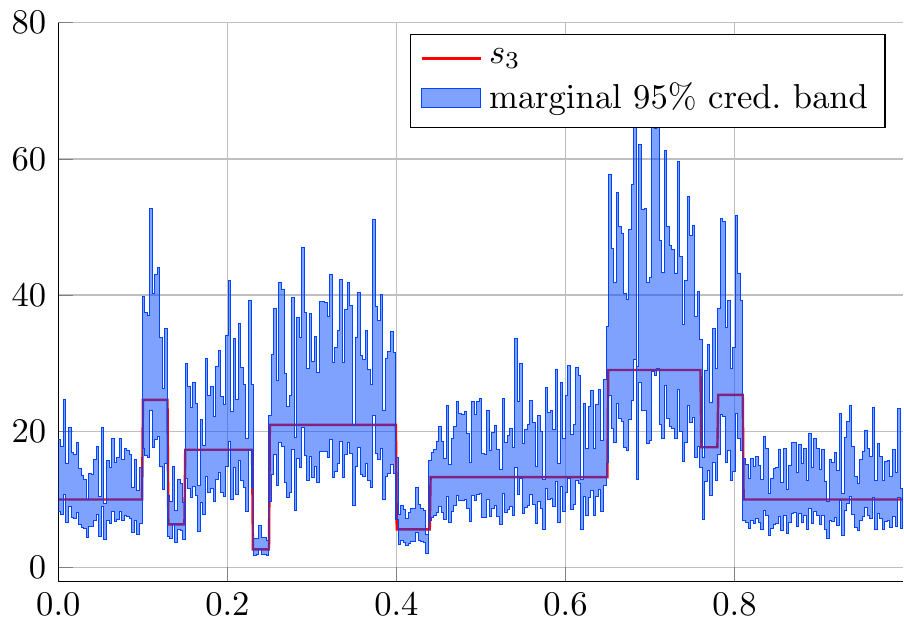}

\bigskip

\includegraphics[width=0.48\textwidth]{\loc 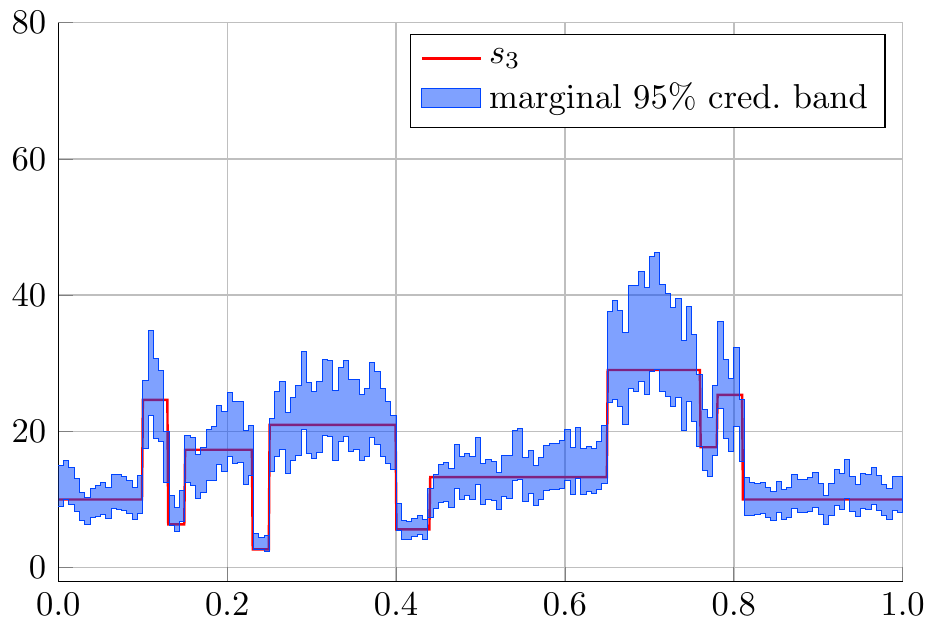}
\includegraphics[width=0.48\textwidth]{\loc 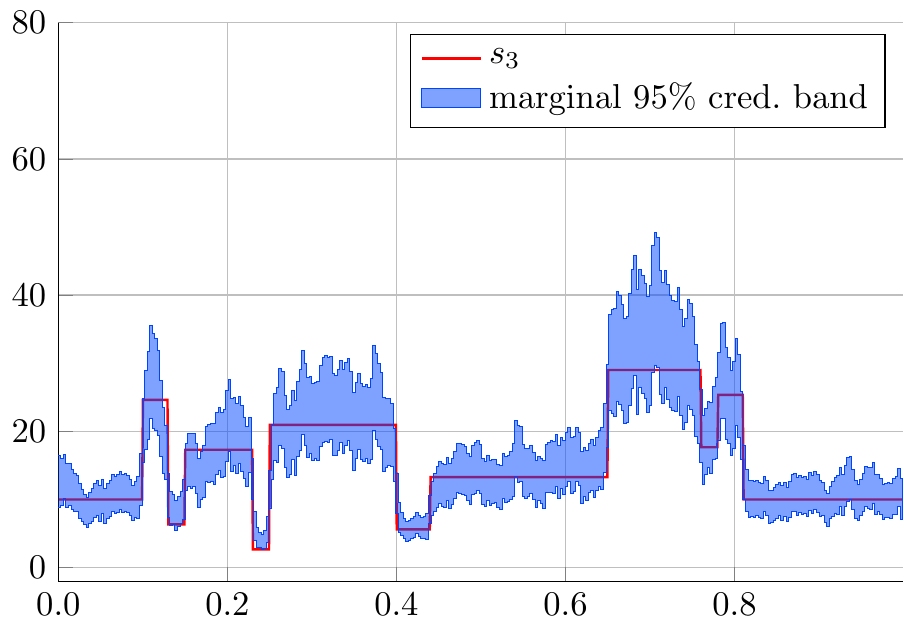}
\caption{Volatility function $s$ with superimposed $95\%$ marginal credible band for the IIG prior $\ig(0.1,0.1)$, using $N=160$ (top left panel) and $N=320$ bins (top right panel). Volatility function $s$ from \eqref{s3} with superimposed $95\%$ marginal credible band for the IGMC prior, using $N=160$ (bottom left panel) and  $N=320$ bins (bottom right panel); in both cases, $\alpha_1=0.1$ and $\alpha =\alpha_{\zeta} \sim\ig(0.3,0.3)$.}
\label{fig:blocks:comp}
\end{figure}

The top left and top right panels of Figure~\ref{fig:blocks:comp} give estimation results obtained with the IIG prior-based approach from \cite{gugu17}. The number of bins was again $N=160$ and $N=320$, and in both these cases we used diffuse independent $\ig(0.1,0.1)$ priors on the coefficients of the (squared) volatility function (see \cite{gugu17} for details). These results have to be contrasted to those obtained with the IGMC prior, plotted in the bottom left and bottom right panels of Figure~\ref{fig:blocks:comp}, where we assumed $\alpha_1=0.1$ and $\alpha =\alpha_{\zeta} \sim\ig(0.3,0.3)$. The following conclusions emerge from Figure~\ref{fig:blocks:comp}:
\begin{itemize}
\item Although both the IGMC and IIG approaches recover globally the shape of the volatility function, the IIG approach results in much greater uncertainty in inferential conclusions, as reflected in wider marginal confidence bands. The effect is especially pronounced in the case $N=320$, where the width of the band for the IIG prior renders it almost useless for inference.
\item The bands based on the IGMC prior look more `regular' than the ones for the IIG prior.
\item Comparing the results to Figure~\ref{fig:blocks}, we see the benefits of equipping the hyperparameters $\alpha,\alpha_{\zeta}$ with a prior: credible bands in Figure~\ref{fig:blocks} do not adequately capture two dips of the function $s$ right before and after the point $t=0.2$, since $s$ completely falls outside the credible bands there. Thus, an incorrect amount of smoothing is used in Figure~\ref{fig:blocks}.
\item The method based on the IIG prior is sensitive to the bin number selection: compare the top left panel of Figure~\ref{fig:blocks:comp} using $N=160$ bins to the top right panel using $N=320$ bins, where the credible band is much wider. On the other hand, the method based on the IGMC prior automatically rebalances the amount of smoothing it uses with different numbers of bins $N$, thanks to the hyperprior on the parameters $\alpha,\alpha_{\zeta}$; in fact, the bottom two plots in Figure~\ref{fig:blocks:comp} look similar to each other.
\end{itemize}

\subsection{CIR model}

Our core estimation procedure, as described in the previous sections, assumes that the volatility function is deterministic. In this subsection, however, in order to test the limits of applicability of our method and possibilities for future extensions, we applied it to a case where the volatility function was stochastic. The study in \cite{mykland12} lends support to this approach, but here we concentrate on practical aspects and defer the corresponding theoretical investigation until another occasion.

Specifically, we considered the Cox-Ross-Ingersoll (CIR) model or the square root process,
\begin{equation}\label{cir}
\D  X_t = (\eta_1-\eta_2X_t)\D  t+\eta_3\sqrt{X_t}\D  W_t, \quad X_0=x>0, \quad t\in [0,T].
\end{equation}
Here $\eta_1,\eta_2,\eta_3>0$ are parameters of the model. This diffusion process was introduced in \cite{feller51a} and \cite{feller51b}, and gained popularity in finance as a model for short-term interest rates, see \cite{cox85}. The condition $2\eta_1 > \eta_3^2$ ensures strict positivity and ergodicity of $X$. The volatility function $s_0$ from \eqref{sde} now corresponds to a realisation of a stochastic process $t\mapsto \eta_3 \sqrt{X_t}$, where $X$ solves the CIR equation \eqref{cir}.

\begin{figure}
\includegraphics[width=0.48\textwidth]{\loc 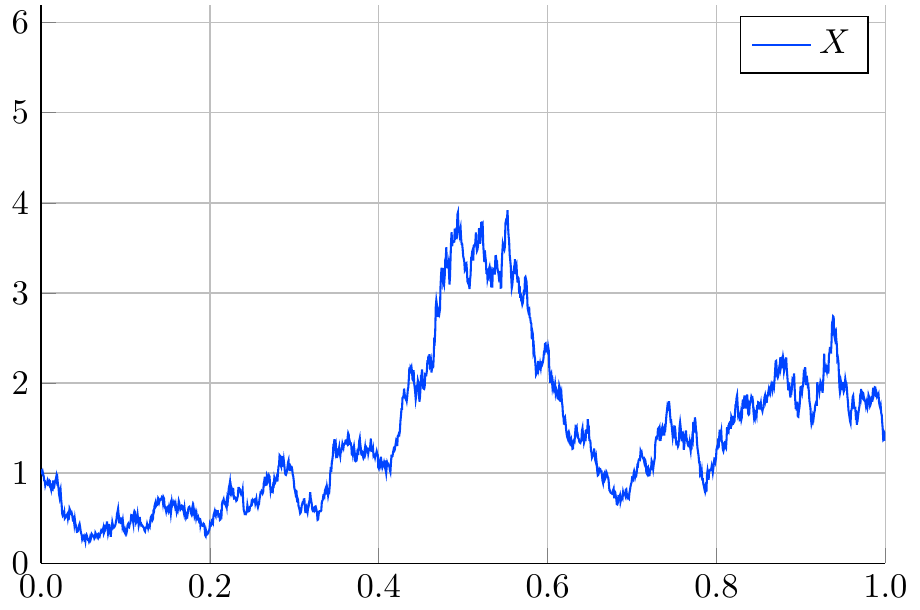}
\includegraphics[width=0.48\textwidth]{\loc 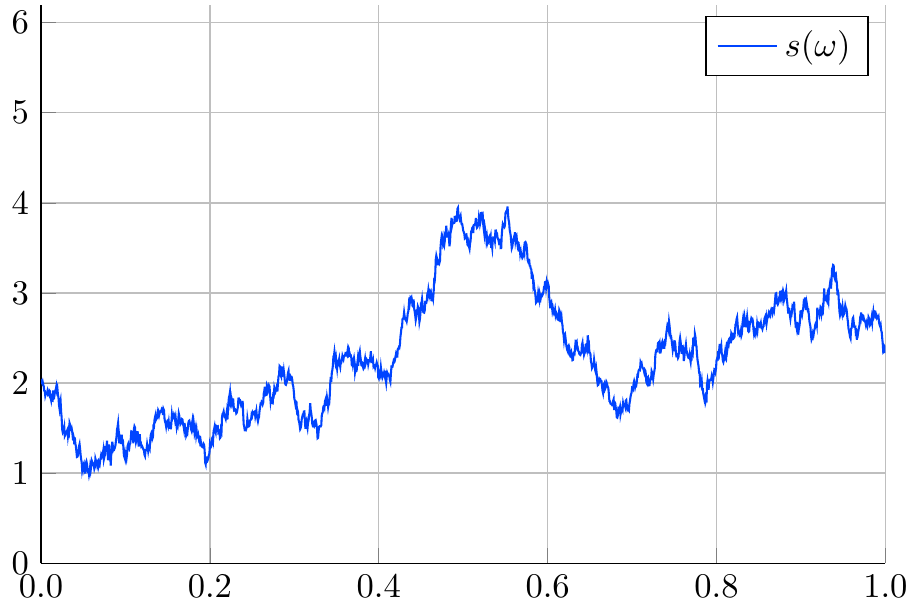}
\caption{The  sample path of the process $X$ from \eqref{cir} (left panel) and the corresponding realised volatility function $s(\omega)$  (right panel). The parameter values are given in \eqref{cir:params}.}
\label{fig:cirpath}
\end{figure}

\begin{figure}
\includegraphics[width=0.48\textwidth]{\loc 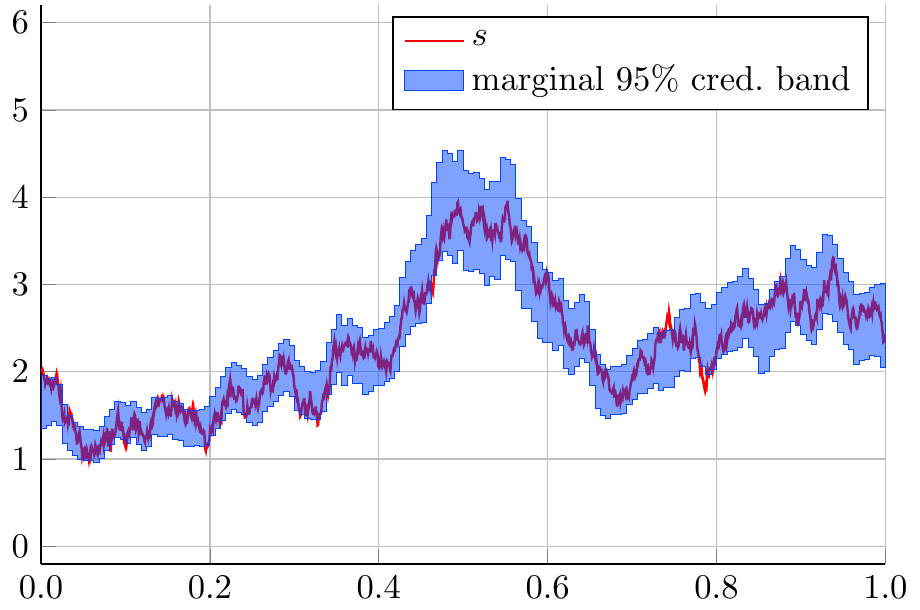}
\includegraphics[width=0.48\textwidth]{\loc 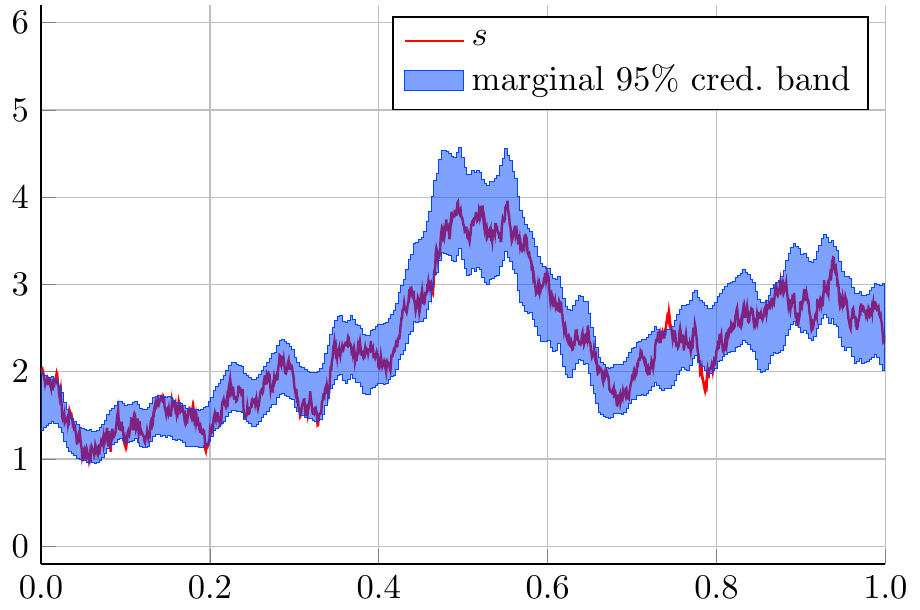}
\caption{Volatility function $s$ from \eqref{cir} with superimposed $95\%$ marginal credible band for the IGMC prior, using $N=160$ (left panel) and  $N=320$ bins (right panel); in both cases, $\alpha_1=0.1$ and $\alpha =\alpha_{\zeta} \sim\ig(0.3,0.3)$.}
\label{fig:cir:comp}
\end{figure}

We took arbitrary parameter values
\begin{equation}
\label{cir:params}
\eta_1=6, \quad \eta_2=3, \quad \eta_1=2, \quad x=1.
\end{equation}
A sample path of $X$ is plotted in the left panel of Figure~\ref{fig:cirpath}, whereas the corresponding volatility is given in the right panel of the same figure. In Figure~\ref{fig:cir:comp} we display estimation results obtained with the IGMC prior, using $N=160$ and $N=320$ bins and hyperparameter specifications $\alpha_1=0.1$ and $\alpha =\alpha_{\zeta} \sim\ig(0.3,0.3)$. A conclusion that emerges from this figure is that our Bayesian method captures the overall shape of the realised volatility in a rather satisfactory manner.

\section{Dow-Jones industrial averages}
\label{section:realdata}

\begin{figure}
\includegraphics[width=0.48\textwidth]{\loc 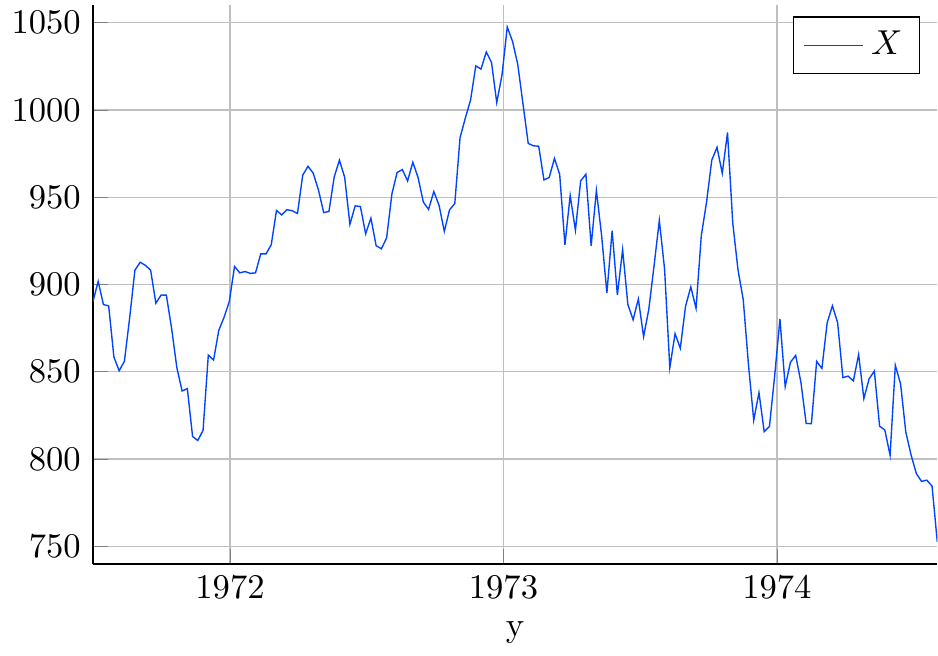}
\includegraphics[width=0.48\textwidth]{\loc 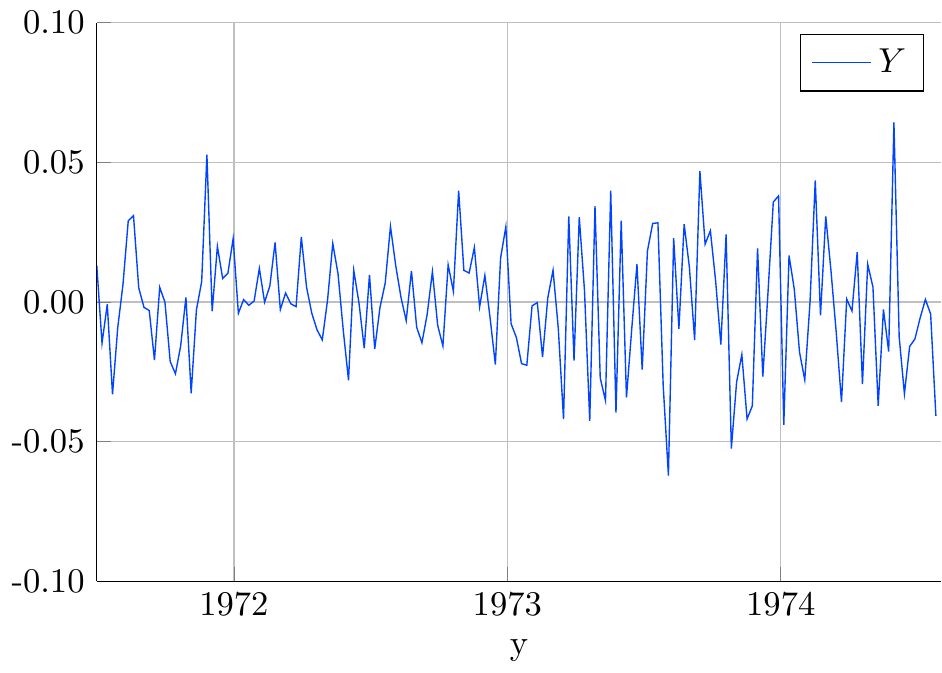}
\caption{Dow-Jones weekly closings of industrial averages over the period 2 July 1971 -- 2 August 1974 (left panel) and the corresponding returns (right panel).}
\label{fig:dwj}
\end{figure}

In this section we provide a reanalysis of a classical dataset in change-point detection in time series; see, e.g., \cite{chen12},  \cite{diaz82}, \cite{hsu77}, \cite{hsu79} and \cite{iacus08}. Specifically, we consider weekly closing values of the Dow-Jones industrial averages in the period 2 July 1971 -- 2 August 1974. In total there are 162 observations available, which constitute a relatively small sample, and thus the inference problem is rather nontrivial. The data can be accessed as the dataset {\tt{DWJ}} in the {\tt{sde}} package (see \cite{iacus16}) in {\bf{R}} (see \cite{r}). See the left panel of Figure~\ref{fig:dwj} for a visualisation. In \cite{iacus08} the weekly data $X_{t_i}$, $i=1,\ldots,n$, are transformed into returns $Y_{t_i}=(X_{t_i}-X_{t_{i-1}})/X_{t_{i-1}}$, and the least squares change-point estimation procedure from \cite{degregorio08} has been performed. Reproducing the corresponding computer code in {\bf R} results in a change-point estimate of 16 March 1973. That author speculates that this change-point is related to the Watergate scandal.

Similar to \cite{iacus08}, parametric change-point analyses in  \cite{chen12},  \cite{diaz82} and \cite{hsu79} give a change-point in the third week of March 1973. However, as noted in \cite{iacus08}, examination of the plot of the time series $Y_{t_i}$ (see Figure \ref{fig:dwj}, the right panel) indicates that another change-point may be present in the data. Then dropping observations after 16 March 1973 and analysing the data for existence of a change-point using only the initial segment of the time series, the author discovers another change-point on 17 December 1971, which he associates with suspending the convertibility of the US dollar into gold under President Richard Nixon's administration.

\begin{figure}
\includegraphics[width=0.48\textwidth]{\loc 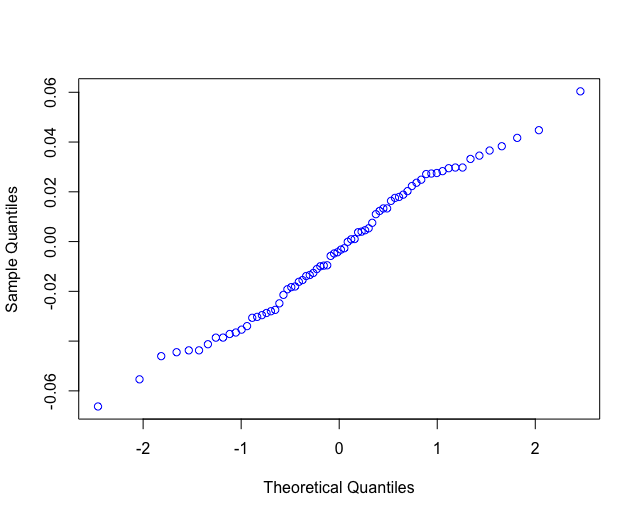}
\includegraphics[width=0.48\textwidth]{\loc 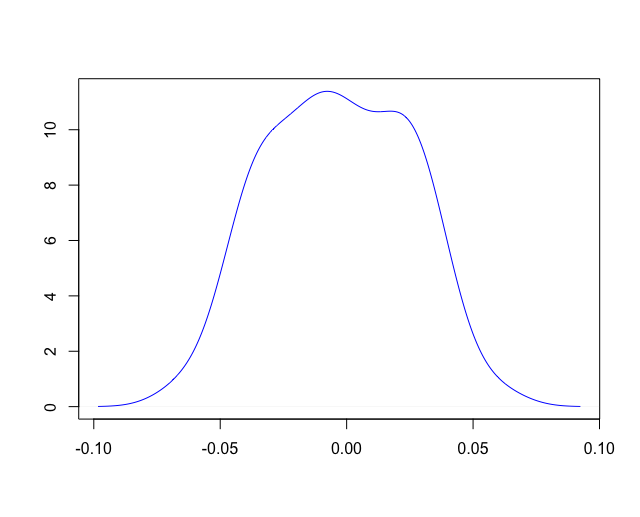}
\caption{QQ plot of the returns of Dow-Jones weekly closings of industrial averages over the period 16 March 1973 -- 2 August 1974 (left panel) and a kernel density estimate of the same data (right panel).}
\label{fig:qqplot}
\end{figure}

From the above discussion it should be clear that nonparametric modelling of the volatility may provide additional insights for this dataset. We first informally investigated the fact whether an SDE driven by the Wiener process is a suitable model for the data at hand. Many of such models, e.g.\ the geometric Brownian motion, a classical model for evolution of asset prices over time (also referred to as the Samuelson or Black-Scholes model), rely on an old tenet that returns of asset prices follow a normal distribution. Although the assumption has been empirically disproved for high-frequency financial data (daily or intraday data; see, e.g.,  \cite{carr02}, \cite{eberlein95} and \cite{kuchler91}), its violation is less severe for widely spaced data in time (e.g.\ weekly data, as in our case). In fact, the Shapiro-Wilk test that we performed in {\bf R} on the returns past the change-point 16 March 1973 did not reject the null hypothesis of normality ($p$-value 0.4). On the other hand, the quantile-quantile (QQ) plot of the same data does perhaps give an indication of a certain mild deviation from normality, see Figure~\ref{fig:qqplot}, where we also plotted a kernel density estimate of the data (obtained via the command {\tt density} in {\bf R}, with bandwidth determined automatically through cross-validation).

\begin{figure}
\includegraphics[width=0.48\textwidth]{\loc 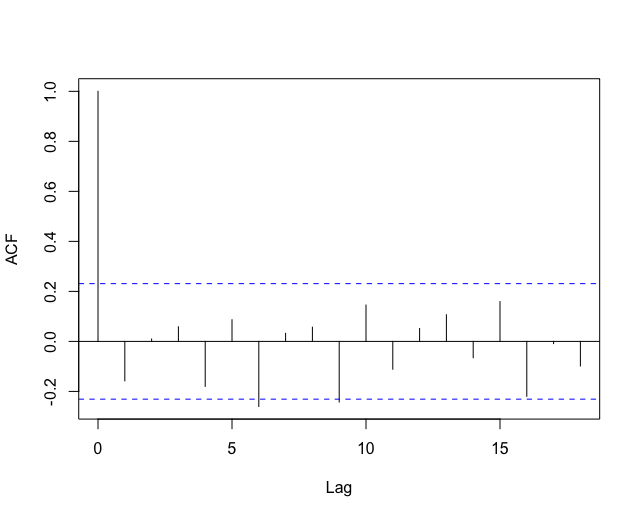}
\includegraphics[width=0.48\textwidth]{\loc 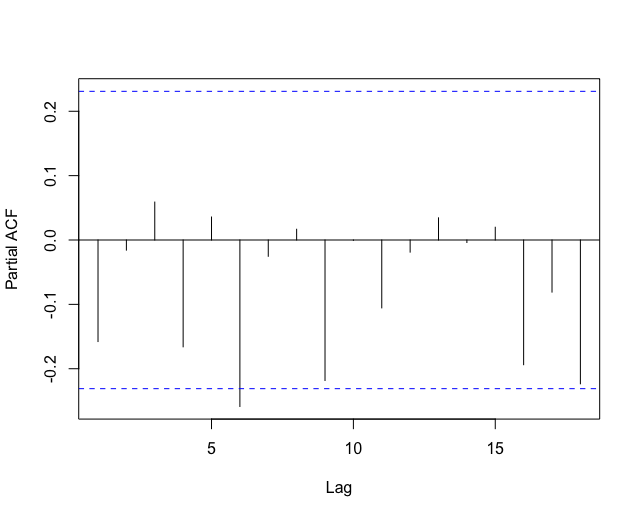}
\caption{Sample autocorrelation (left panel) and partial autocorrelation functions of the returns of Dow-Jones weekly closings of industrial averages over the period 16 March 1973 -- 2 August 1974.}
\label{fig:acf:pacf}
\end{figure}

In Figure~\ref{fig:acf:pacf} we plot the sample autocorrelation and partial autocorrelation functions based on returns $Y_{t_i}$'s past the change-point 16 March 1973. These do not give decisive evidence against the assumption of independence of $Y_{t_i}$'s. Neither does the Ljung-Box test (the test is implemented in {\bf R} via the command {\tt{Box.test}}), which yields a $p$-value $0.057$ when applied with $10$ lags (the $p$-value is certainly small, but not overwhelmingly so).

Summarising our findings, we detected only a mild evidence against the assumption that the returns of the Dow-Jones weekly closings of industrial averages (over the period 16 March 1973 -- 2 August 1974, but similar conclusions can be reached also over the other subperiods covered by the {\tt{DWJ}} dataset) are approximately independent and follow a normal distribution. Thus there is no strong a priori reason to believe that a geometric Brownian motion is an outright unsuitable model in this setting: it can be used as a first approximation. To account for time-variability of volatility (as suggested by the change-point analysis), we incorporate a time-dependent volatility function in the model, and for additional modelling flexibility we also allow a state-dependent drift. Setting $Z_t=\log(X_t/X_0)$, our model is thus given by
\begin{equation}
\label{eq:gbm}
\D  Z_t=b_0(t,Z_t)\D  t + s_0(t)\D  W_t, \quad Z_0=0.
\end{equation}

An alternative here is to directly (i.e.\ without any preliminary transformation) model the Dow-Jones data using equation \eqref{sde}. We consider both possibilities, starting with the model \eqref{eq:gbm}.

\begin{figure}
\includegraphics[width=0.48\textwidth]{\loc 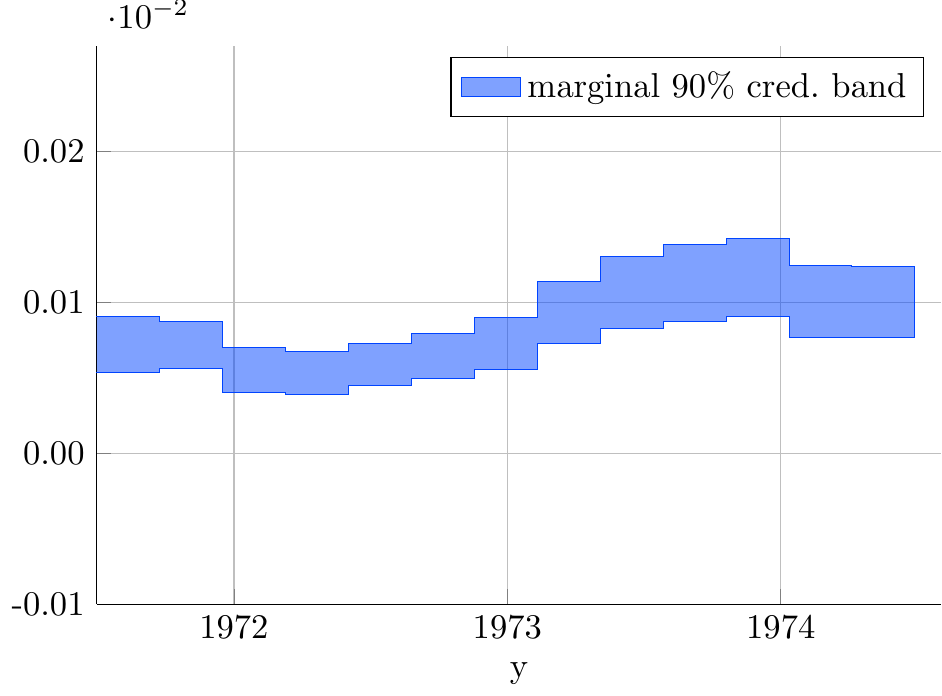}
\includegraphics[width=0.48\textwidth]{\loc 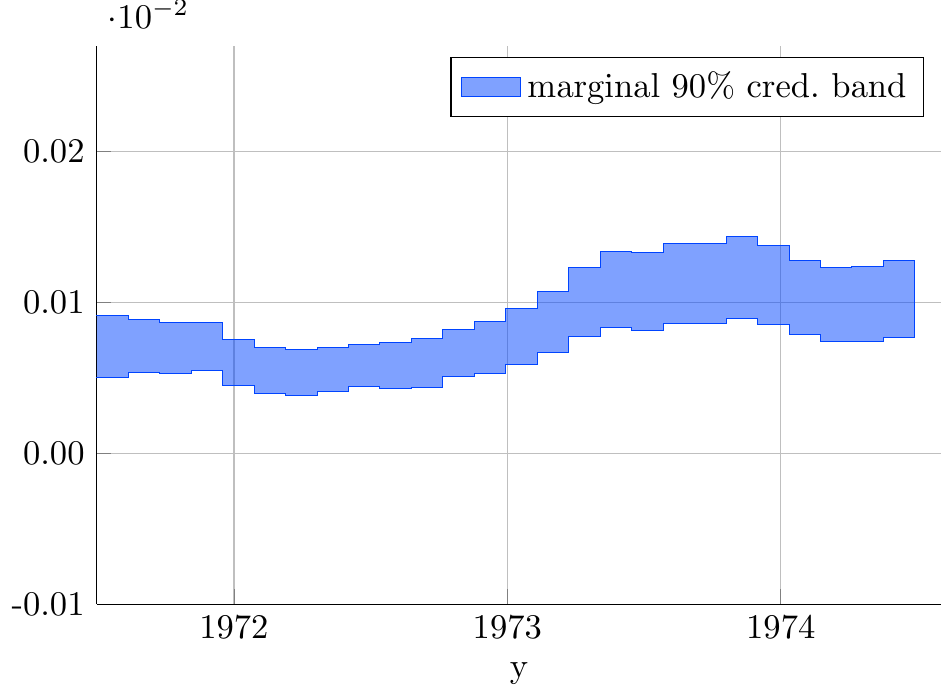}
\caption{Marginal $90\%$ credible bands for the volatility function of the log Dow-Jones industrial averages data. The left panel corresponds to $N=13$ bins, while the right panel to $N=26$ bins.}
\label{fig:logdwj:bayes}
\end{figure}

We used a vague prior on $\theta_1$ corresponding to the limit $\alpha_1\rightarrow 0$, whereas for the other two hyperparameters we assumed $\alpha=\alpha_{\zeta}\sim\ig(0.3,0.3)$. The scaling in the independent Gaussian random walk proposal in the Metropolis-Hastings step was chosen in such a way so as to yield an acceptance rate of ca.~$50\%$. The Gibbs sampler was run for $200 \, 000$ iterations, and the first $1000$ samples were dropped as a burn-in. We present the estimation results we obtained using $N=13$ and $N=26$ bins, see Figure~\ref{fig:logdwj:bayes}. Although the sample size $n$ is quite small in this example, the data are informative enough to yield nontrivial inferential conclusions even with diffuse priors. Both plots in Figure~\ref{fig:logdwj:bayes} are qualitatively similar and suggest:
\begin{itemize}
	\item A decrease in volatility at the end of 1971, which can be taken as corresponding to the change-point in December 1971 identified in \cite{iacus08}. Unlike that author, we do not directly associate it with suspending the convertibility of the US dollar into gold (that took place in August 1971 rather than December 1971).
	\item A gradual increase in volatility over the subsequent period stretching until the end of 1973. Rather than only the Watergate scandal (and a change-point in March 1973 as in \cite{iacus08}), there could be further economic causes for that, such as the 1973 oil crisis and the 1973--1974 stock market crash.
	\item A decrease in volatility starting in early 1974, compared to the immediately preceding period.
\end{itemize}
In general, in this work we do not aim at identifying causes for changes in volatility regimes, but prefer to present our inference results, that may subsequently be used in econometric analyses.

\begin{figure}
\includegraphics[width=0.48\textwidth]{\loc 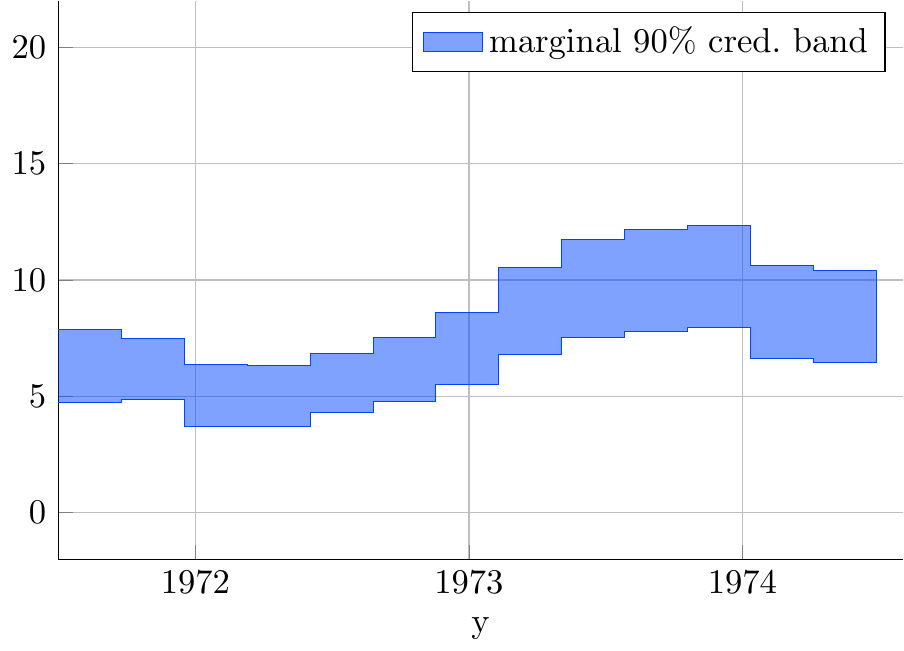}
\includegraphics[width=0.48\textwidth]{\loc 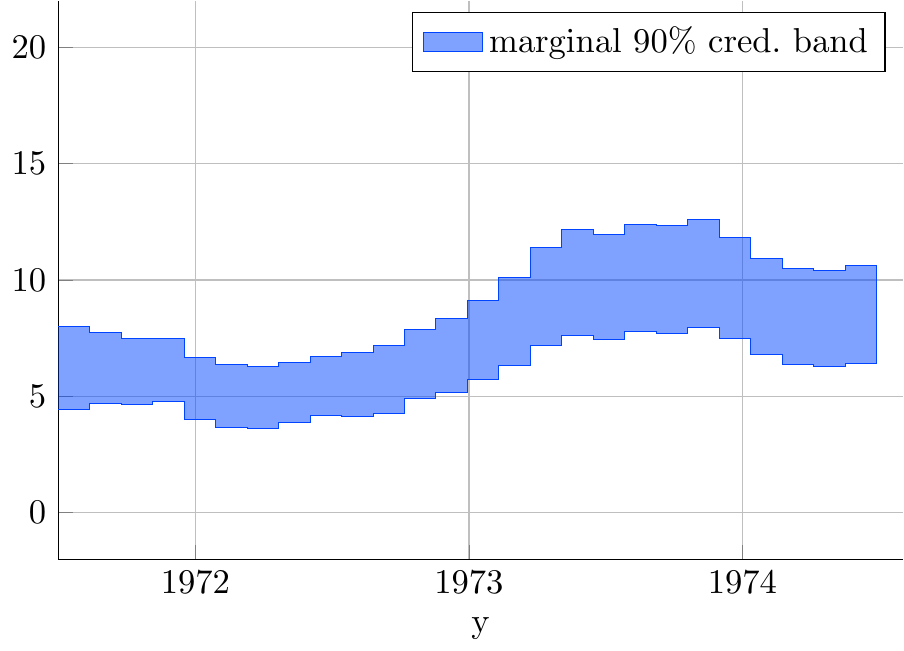}
\caption{Marginal $90\%$ credible bands for the volatility function of the Dow-Jones industrial averages data. The left panel corresponds to $N=13$ bins, while the right panel to $N=26$ bins.}
\label{fig:dwj:bayes}
\end{figure}

Now we move to the Bayesian analysis of the data using model \eqref{sde}. The prior settings were as in the previous case, and we display the results in Figure~\ref{fig:dwj:bayes}. The overall shapes of the inferred volatility functions are the same in both Figure~\ref{fig:logdwj:bayes} and Figure~\ref{fig:dwj:bayes}, and hence similar conclusions apply.

Finally, we stress the fact that our nonparametric Bayesian approach and change-point estimation are different in their scope: whereas our method aims at estimation of the entire volatility function, change-point estimation (as its name actually suggests) concentrates on identifying change-points in the variance of the observed time series, which is a particular feature of the volatility. To that end it assumes the (true) volatility function is piecewise constant, which on the other hand is not an assumption required in our method. Both techniques are useful, and each can provide insights that may be difficult to obtain from another.

%%%%%%%%%%%%%%%%
\section{Conclusions}
\label{section:conclusions}

Bayesian inference for SDEs from discrete-time observations is a difficult task, owing to intractability of the likelihood and the fact that the posterior is not available in closed form. Posterior inference therefore typically requires the use of intricate MCMC samplers. Designing algorithms that result in Markov chains that mix well and explore efficiently the posterior surface is a highly nontrivial problem. Inspired by some ideas from the audio signal processing literature and our earlier work \cite{gugu17}, in this paper we introduced a novel nonparametric Bayesian approach to estimation of the volatility coefficient of an SDE. Our method is easy to understand and straightforward to implement via Gibbs sampling, and performs well in practice. Thereby our hope is that our work will contribute to further dissemination and popularisation of a nonparametric Bayesian approach to inference in SDEs, specifically with financial applications in mind. In that respect, see \cite{gugu18:micro}, that builds upon the present paper and deals with Bayesian volatility estimation under market microstructure noise. Our work can also be viewed as a partial fulfillment of anticipation in \cite{godsill07} that some ideas developed originally in the context of audio and music processing ``will also find use in other areas of science and engineering, such as financial or biomedical data analysis''.

As a final remark, we do not attempt to provide a theoretical, i.e.\ asymptotic frequentist analysis of our new approach here (see, e.g., the recent monograph \cite{ghosal17}, and specifically \cite{gugu17} for such an analysis in the SDE setting), but leave this as a topic of future research.

\section{Formulae for parameter updates}
\label{section:details}

In this section we present additional details on the derivation of the update formulae for the Gibbs sampler from Section~\ref{section:model}. The starting point is to employ the Markov property from \eqref{formula:prior}, and using the standard Bayesian notation, to write the joint density of $\{\zeta_k\}$ and $\{\theta_k\}$ as
\begin{equation}
\label{formula:parameters}
p(\theta_1) \prod_{k=2}^N p(\zeta_{k} | \theta_{k-1}) p(\theta_k | \zeta_k) .
\end{equation}

\subsection{Full conditional distributions} We first indicate how \eqref{fullcond1} was derived. Insert expressions for the individual terms in \eqref{formula:parameters} from \eqref{formula:prior} and collect separately terms that depend on $\theta_k$ only, to see that the density of the full conditional distribution of $\theta_k$, $k=2,\ldots,N-1$, is proportional to
\[
\theta_k^{-\alpha-1} e^{-\alpha/(\theta_k\zeta_k)} \theta_k^{-\alpha_{\zeta}} e^{-\alpha_{\zeta}/(\theta_k\zeta_{k+1})}.
\]
Upon normalisation, this expression is the density of the $\ig(\alpha+\alpha_{\zeta},\alpha \zeta_k^{-1}+\alpha_{\zeta} \zeta_{k+1}^{-1})$ distribution, which proves formula \eqref{fullcond1}. Formula \eqref{fullcondend} follows directly from the last expression in \eqref{formula:prior}. Formula \eqref{fullcond2} is proved analogously to \eqref{fullcond1}. Finally, \eqref{fullcondstart} follows from \eqref{formula:prior} and Bayes' formula. Cf.\ also \cite{dikmen08}, Appendix~B.6.

\subsection{Metropolis-within-Gibbs step}
\label{section:mwg}

Now we describe the Metropolis-Hastings step within the Gibbs sampler, that is used to update the hyperparameters of our algorithm, in case the latter are equipped with a prior. For simplicity, assume $\alpha=\alpha_{\zeta}$ (we note that such a choice is used in practical examples in \cite{cemgil:proc:08}), and suppose $\alpha$ is equipped with a prior, $\alpha\sim\pi$. Let the hyperparameter $\alpha_1$ be fixed. Obviously, $\alpha_{1}$ could have been equipped with a prior as well, but this would have further slowed down our estimation procedure, whereas the practical results in Sections~\ref{section:simulation} and \ref{section:realdata} we obtained are already satisfactory with $\alpha_{1}$ fixed. Using \eqref{formula:prior} and \eqref{formula:parameters}, one sees that the joint density of $\{\zeta_k\}$, $\{\theta_k\}$ and $\alpha$ is proportional to
\begin{multline*}
\pi(\alpha) \times \theta_1^{-\alpha_1 - 1} \times e^{-\alpha_1\theta_1^{-1}} \\
 \times \prod_{k=2}^N \left\{ \frac{\alpha^{\alpha}}{\Gamma(\alpha) \theta_{k-1}^{\alpha} } \zeta_k^{-\alpha-1} e^{-\alpha/(\theta_{k-1}\zeta_k)} \frac{\alpha^{\alpha}}{\Gamma(\alpha) \zeta_k^{\alpha}} \theta_{k}^{-\alpha-1} e^{-\alpha/(\theta_k\zeta_{k})} \right\}.
\end{multline*}
This in turn is proportional to
\begin{multline*}
q(\alpha) = \pi(\alpha) \times\left(\frac{\alpha^{\alpha}}{\Gamma(\alpha)}\right)^{2(N-1)}  \times \prod_{k=2}^N (\theta_{k-1} \theta_k \zeta_k^2)^{-\alpha} \\
 \times \exp\left(-\alpha\sum_{k=2}^N\frac{1}{\zeta_k}\left( \frac{1}{\theta_{k-1}} + \frac{1}{\theta_k}\right)\right).
\end{multline*}
The latter expression is an unnormalised full conditional density of $\alpha$, and can be used in the Metropolis-within-Gibbs step to update $\alpha$. 

The rest of the Metropolis-Hastings step is standard, and the following approach was used in our practical examples: pick a proposal kernel $g(\alpha^{\prime} \mid \alpha)$, for instance a Gaussian random walk proposal $g(\alpha^{\prime} \mid \alpha)=\phi_{\sigma}(\alpha^{\prime} - \alpha)$, where $\phi_{\sigma}$ is the density of a normal random variable with mean zero and variance $\sigma^2$. Note that this specific choice may result in proposing a negative value $\alpha^{\prime}$, which needs to be rejected straightaway as invalid. Then, for computational efficiency, instead of moving to another step within the Gibbs sampler, one keeps on proposing new values $\alpha^{\prime}$ until a positive one is proposed. This is then accepted with probability
\[
A=\min\left(1,\frac{q(\alpha^{\prime})}{q(\alpha)} \frac{\Phi_{\sigma}(\alpha)}{\Phi_{\sigma}(\alpha^{\prime})}\right),
\]
where $\Phi_{\sigma}(\cdot)$ is the cumulative distribution function of a normal random variable with mean zero and variance $\sigma^2$; otherwise the current value $\alpha$ is retained. See \cite{wilkinson12} for additional details and derivations. Finally, one moves to other steps in the Gibbs sampler, namely to updating $\zeta_k$'s and $\theta_k$'s, and iterates the procedure. The acceptance rate in the Metropolis-Hastings step can be controlled through the scale parameter $\sigma$ of the proposal density $\phi_{\sigma}$. Some practical rules for determination of an optimal acceptance rate in the Metropolis-Hastings algorithm are given in \cite{gelman:proc:96}, and those for the Metropolis-within-Gibbs algorithm in \cite{sherlock10}.

\bibliography{bibliography}

\end{document}